\documentclass[11pt, a4paper, UKEnglish]{article}
\usepackage{a4wide}
\usepackage{latexsym}               %latex-spezifische Symbole                          %%%
\usepackage{amsfonts}
\usepackage{mathrsfs}
\usepackage[ansinew]{inputenc}      %deutsche Umlaute, "" usw                          %%%
\usepackage{amssymb}                %besondere Symbole                                  %%%
\usepackage{amsmath}                %besondere Symbole                                  %%%
\usepackage[]{graphicx}
\usepackage{floatflt}
\usepackage{mathrsfs}
\usepackage{verbatim}
\usepackage{fancyhdr}
\usepackage[font=scriptsize,labelfont=bf,skip=6pt]{caption}
\usepackage[T1]{fontenc}
\usepackage{subcaption}
\usepackage{color}
\usepackage{overpic}
\usepackage{xspace}
\usepackage{titlesec}
\usepackage{lineno}
\usepackage{changepage}
\usepackage{enumitem}
\usepackage{setspace}%\onehalfspacing
\usepackage{array}
\usepackage{titlesec}
\usepackage{rotating}
\usepackage{eurosym}
\usepackage{footnote}
\usepackage{tcolorbox}
\usepackage{authblk}
\definecolor{myviolet}{rgb}{0.73,0.56,0.64}
\definecolor{myblue}{rgb}{0.00,0.09,0.73}
\definecolor{myorange}{rgb}{0.99,0.88,0.8}
\definecolor{myred}{rgb}{0.64,0.09,0.16}
\usepackage[urlcolor=myblue,   % farbige Hyperlinks
            colorlinks=true, % farbige Querverweise
            linkcolor=myblue,  % Blau als Farbe fuer Links
            citecolor=myred,
            bookmarks,       % Navigationsleiste im Acrobat
            bookmarksnumbered=true]{hyperref}

\usepackage[backend=biber,style=phys,eprint=true,hyperref=true,biblabel=brackets]{biblatex}

\def\ifb{{\ \rm fb}^{-1}}
\def\iab{{\ \rm ab}^{-1}}

\def\metre{\text{m}\xspace}
\def\sqm{\ensuremath{\text{m}^2}\xspace}

\def\mllp{\ensuremath{m_\text{LLP}}\xspace}
\def\meff{\ensuremath{m_\text{med}}\xspace}

\newcommand{\nc}{\newcommand}
\nc{\beq}{\begin{equation}}
\nc{\eeq}{\end{equation}}
\nc{\barray}{\begin{eqnarray}}
\nc{\earray}{\end{eqnarray}}
\nc{\barrayn}{\begin{eqnarray*}}
\nc{\earrayn}{\end{eqnarray*}}
\nc{\bcenter}{\begin{center}}
\nc{\ecenter}{\end{center}}
\nc{\mc}{\mathcal}
\newcommand{\gsim}{\lower.7ex\hbox{$\;\stackrel{\textstyle>}{\sim}\;$}}% This is just \gtrsim
\newcommand{\lsim}{\lower.7ex\hbox{$\;\stackrel{\textstyle<}{\sim}\;$}}%This is just \lesssim

\newcommand{\etmiss}{\ensuremath{E \kern-0.6em\slash_{\rm T}}\xspace}
\newcommand{\etmissx}{\ensuremath{E \kern-0.6em\slash_{\rm x}}\xspace}
\newcommand{\etmissy}{\ensuremath{E \kern-0.6em\slash_{\rm y}}\xspace}

\newcommand{\const}{\operatorname{const}}

\newcommand{\cm}{\ensuremath{\textnormal{cm}}\xspace}

\newcommand{\GeV}{\ensuremath{\textnormal{GeV}}\xspace}

\newcommand{\stat}{\ensuremath{\textnormal{(stat)}}\xspace}
\newcommand{\syst}{\ensuremath{\textnormal{(syst)}}\xspace}

\newcommand{\dif}{\ensuremath{{\rm d}}\xspace}
\newcommand{\dR}{\ensuremath{\Delta R}\xspace}
\newcommand{\leff}{\ensuremath{L_\text{eff}}\xspace}

\newcommand{\met}{\ensuremath{E_T^\text{miss}}\xspace}

\newcommand{\fb}{\ensuremath{{\rm fb}^{-1}}\xspace}

\newcommand{\pt}{\ensuremath{p_{\mathrm{T}}\xspace}}

\newcommand{\br}{\ensuremath{\mathcal Br}\xspace}

\newcommand{\LI}{\ensuremath{\Lambda_I}\xspace}
\newcommand{\LIx}[1]{\ensuremath{\Lambda_{I,\text{#1}}}\xspace}

\newtcolorbox{mybox}{
    arc=0pt,
    boxrule=0pt,
    colback=myorange,
    width=\textwidth,   % this option controls the width of the box
    colupper=black,
}

\usepackage{multirow}

\graphicspath{{figaux/}}

\title{
ANUBIS: Proposal to search for long-lived neutral particles \\
in CERN service shafts
}

\author[1]{Martin Bauer\thanks{martin.m.bauer@durham.ac.uk}}
\affil[1]{Department of Physics, University of Durham}
\author[2]{Oleg Brandt\thanks{oleg.brandt@cern.ch}}
\affil[2]{Cavendish Laboratory, University of Cambridge}
\author[3]{Lawrence Lee\thanks{lawrence.lee.jr@cern.ch}}
\affil[3]{Department of Physics, University of Tennessee, Knoxville}
\author[4]{Christian Ohm\thanks{chohm@kth.se}}
\affil[4]{KTH Royal Institute of Technology and Oskar Klein Centre}

\providecommand{\keywords}[1]{\textbf{\textit{Key Words--}} #1}

\begin{document}

%\thanks{\href{mailto:martin.m.bauer@durham.ac.uk}{martin.m.bauer@durham.ac.uk}}
%\thanks{\href{mailto:oleg.brandt@cern.ch}{oleg.brandt@cern.ch}, corresponding author}
%\thanks{\href{mailto:lawrence.lee.jr@cern.ch}{lawrence.lee.jr@cern.ch}}
%\thanks{\href{mailto:chohm@kth.se}{chohm@kth.se}}

\maketitle

\begin{abstract}
Long-lived particles are predicted by many extensions of the Standard Model and have been gaining interest in recent years.
In this paper the original proposal is presented for AN Underground Belayed In-Shaft (ANUBIS) detector that substantially extends the sensitivity to particle lifetimes by instrumenting the existing service shafts above the ATLAS or CMS experiments with tracking stations. 
For scenarios with electrically neutral long-lived particles with $m \gtrsim 1$~GeV produced at the electroweak scale and above, the lifetime reach is increased by 2-3 orders of magnitude compared to currently operating and approved future experiments at the LHC. 
The original ANUBIS detector design proposal is outlined along with the projected costs.
\end{abstract}

%\vspace{30mm}
\vfill

\setcounter{equation}{0} \setcounter{footnote}{0}
%\pagenumbering{arabic}

\clearpage

%%%%%%%%%%%%%%%%%%%%%%%%%%%%
\section{Introduction}%\label{sec:intro}
%%%%%%%%%%%%%%%%%%%%%%%%%%%%

The Standard Model (SM) is one of the most accurate scientific theories ever devised and has passed decades of experimental tests. Yet, there are many fundamental questions that can not be answered without physics beyond the SM~(BSM). This includes the nature of dark matter~\cite{Hall:2010jx,Cheung:2010gk,Zurek:2013wia}, neutrino masses~\cite{Bondarenko:2018ptm}, the origin of the matter-antimatter asymmetry~\cite{Sakharov:1967dj}, the puzzling unnaturalness of the electroweak scale~\cite{Martin:1997ns}, and  a theory of quantum gravity. 
Many proposed theoretical scenarios that address these questions predict particles with long lifetimes on the scale of typical collider experiments~\cite{Barbier:2004ez,Giudice:1998bp,Han:2007ae,bib:htoss,Chacko:2005pe,Cai:2008au}, see also Refs.~\cite{PBC:2025sny,Antel:2023hkf,Lee:2018pag,Alimena:2019zri} for recent reviews of the LLP landscape. A number of dedicated experiments have been proposed to search for such long-lived particles~(LLPs) with decay lengths $L$ of $\mathcal{O}(10~{\rm m})$ and above. 

New beam-dump experiments such as SHiP~\cite{%Bonivento:2013jag,Alekhin:2015byh,
Ahdida:2654870} can provide orders of magnitudes better sensitivity to their target models compared to their predecessors, provided they are within kinematic reach~\cite{Antel:2023hkf}.  
While SHiP will require dedicated operation with the 400~GeV SPS beam, various proposals to utilize the Large Hadron Collider (LHC) beams without interference with the LHC physics program have been put forward. 
The FASER detector with an active volume of $\sim$1~m$^3$ to register LLP decays was installed 480~m downstream from the ATLAS interaction point~\cite{Feng:2017uoz,Ariga:2018uku} and has been taking data since 2022; MATHUSLA is a proposed large-scale surface detector instrumenting an active volume of $\sim$8$\times 10^4$~m$^3$ above ATLAS or CMS~\cite{Curtin:2018mvb} that was revised to a volume of $\sim$2$\times 10^4$~m$^3$ above the CMS experiment~\cite{MATHUSLA:2025zyt}; %40 m × 40 m × 11 m
and CODEX-b is a proposed $\sim$10$^3$~m$^3$ detector to be installed in the LHCb cavern~\cite{Gligorov:2017nwh}. AL3X proposes to search for LLPs using a cylindrical $\sim$900~m$^3$ detector inside the L3 magnet and the time-projection chamber of the ALICE experiment~\cite{Gligorov:2018vkc}. 
Dedicated experiments for LLPs with exotic electromagnetic charges include MilliQan~\cite{Haas:2014dda}, searching for millicharged particles in the drainage gallery of CMS, and MoEDAL~\cite{Pinfold:2009oia}, looking for highly ionizing particles like magnetic monopoles at LHCb alongside MAPP~\cite{Pinfold:2675883}. Timing information can be explored at ATLAS and CMS~\cite{Sirunyan:2019gut,Liu:2018wte}.

Among the proposals, {\em forward} collider experiments like FASER and  beam-dump experiments like SHiP target relatively light LLPs with invariant masses of $\mathcal{O}(1~\GeV)$ produced at relatively low partonic centre-of-mass energies of $\sqrt{\hat s} \lesssim10$~GeV due to kinematic constraints.
{\em Transverse} collider experiments such as MATHUSLA are imperative to search for LLPs with $L \gsim10~$m produced at the electroweak scale or above, i.e., $\sqrt{\hat s}\gtrsim 80$~GeV~\cite{Chou:2016lxi, Curtin:2018mvb}. 
Such scenarios cannot be probed at forward experiments nor at main LHC detectors ATLAS, CMS, or LHCb. 

\enlargethispage{5mm}

\begin{figure}[h]
\begin{center}
\includegraphics[width=0.35\textwidth]{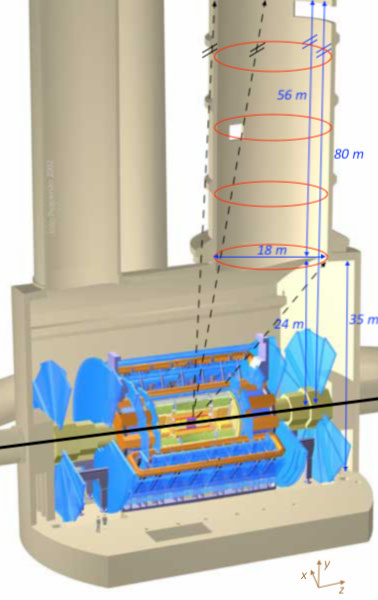}
\end{center}
\caption{
Schematic view of the UX15 cavern with the ATLAS detector and the access shafts PX16, PM15, and PX14 (from left to right). The beamline is indicated by the solid thick line, while the dimensions of the cavern and the main shaft are shown by the arrows. Potential positions of the ANUBIS tracking stations spaced about 18.5~m apart along the 18~m wide PX14 shaft are indicated by the ellipses. Representative trajectories of LLPs falling within the ANUBIS acceptance are indicated by dashed lines. Figure adapted from Ref.~\cite{Aad:2008zzm}.
}
\label{fig:cavern}
\end{figure}

This paper documents the original proposal of a new {\em transverse} experiment to significantly extend the growing LLP search programme by taking advantage of the 18~m diameter, 56~m long PX14 installation shaft of the ATLAS experiment. 
The detector structure, \emph{AN Underground Belayed In-Shaft search experiment} (ANUBIS), would consist of four tracking stations belayed into the shaft and affixed to its walls, instrumenting an active volume of about 15,000~m$^3$ with dedicated LLP detectors, as illustrated in Fig.~\ref{fig:cavern}.
Using the existing infrastructure of the PX14 shaft yields acceptance for a wide range of LLP lifetimes while maintaining competitive installation and operation costs. 
This proof-of-concept study makes a concrete design proposal for ANUBIS, discusses the possible detector technologies,  provides first approximate cost estimates, and compares the expected sensitivity reach to that of the ATLAS and the proposed CODEX-b and MATHUSLA experiments for a benchmark model with exotic Higgs boson decays into LLPs. 
The benchmark, motivated by Neutral Naturalness~\cite{Giudice:1998bp,Burdman:2006tz,Argyropoulos:2021sav}, is endorsed by the Physics Beyond Colliders initiative~\cite{Beacham:2019nyx} and featured in its ESPPU 2026 submission~\cite{PBC:2025sny}.
%that cannot be probed at forward or beam dump experiments.

%%%%%%%%%%%%%%%%%%%%%%%%%%%%
\section{The ANUBIS Detector Concept}%
\label{sec:concept}
%%%%%%%%%%%%%%%%%%%%%%%%%%%%

\begin{figure}[tb]
\centering
\begin{subfigure}{0.49\textwidth}
\includegraphics[height=0.8\textwidth]{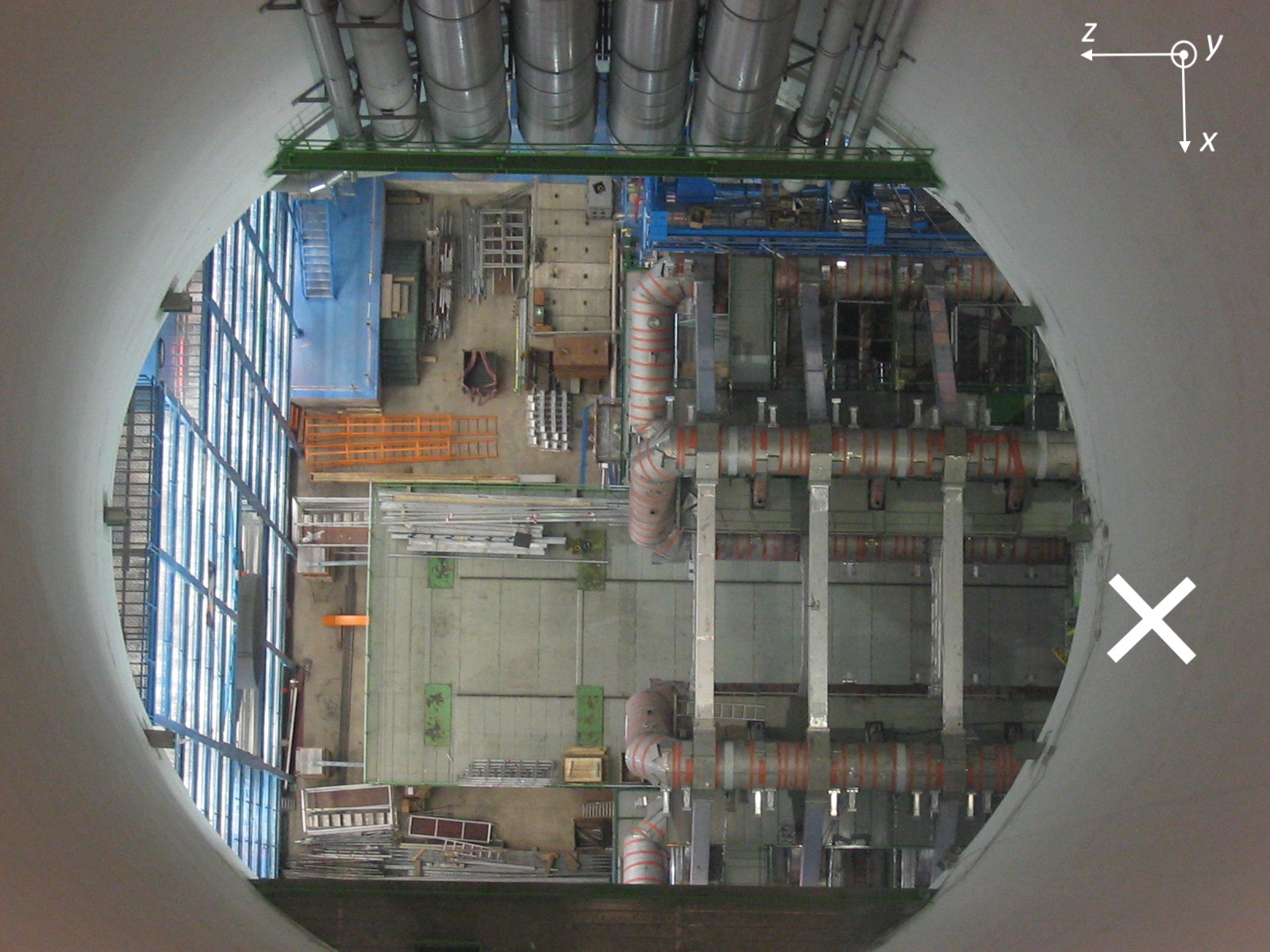}
\caption{}
\label{fig:shaft_above}
\end{subfigure}
\begin{subfigure}{0.49\textwidth}
\includegraphics[height=0.8\textwidth]{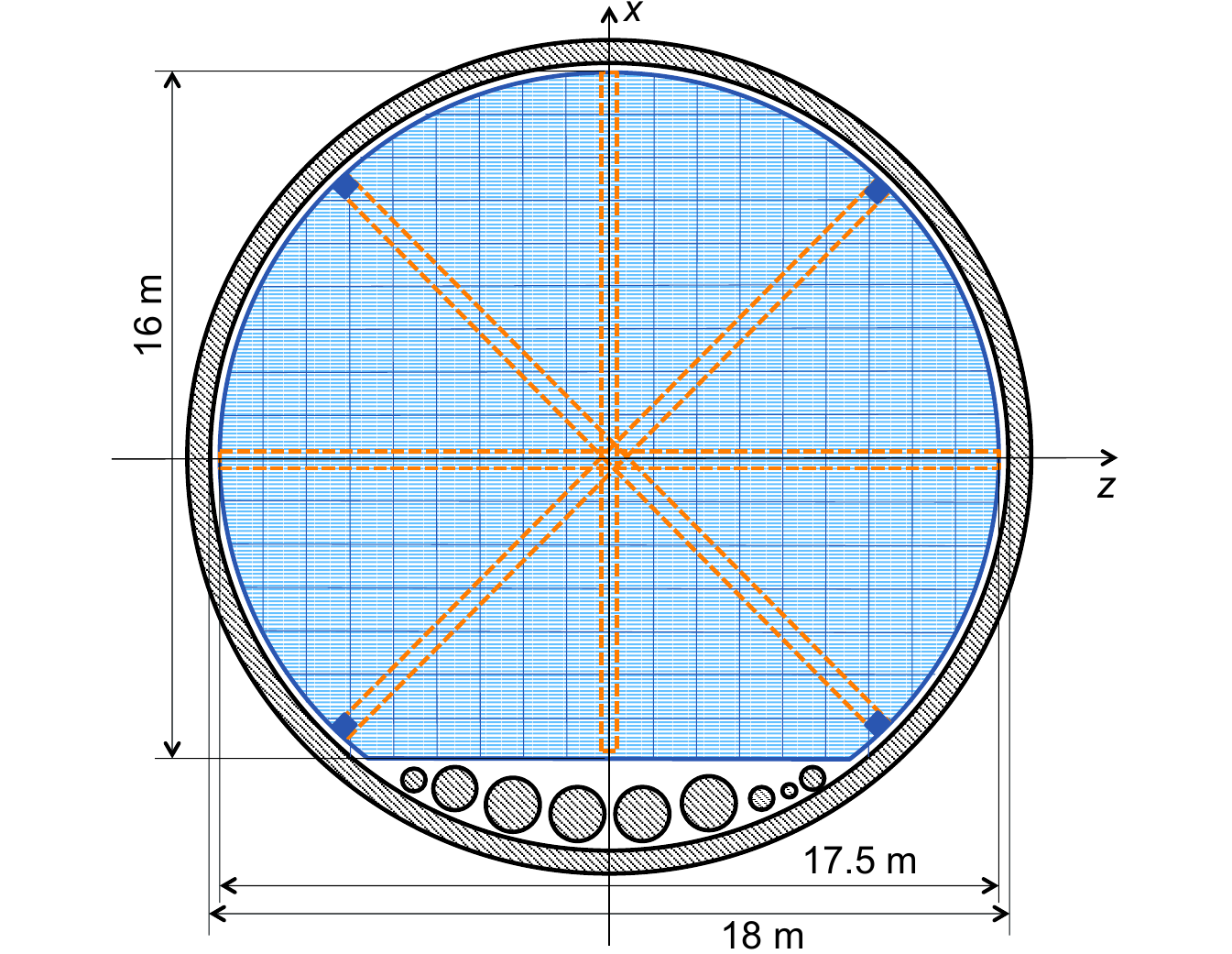}
\caption{}
\label{fig:anubis_ts}
\end{subfigure}
\caption{
\label{fig:shaft}
\textbf{(a)}
View from the surface down the 92~m deep and 18~m wide main shaft PX14 of the ATLAS Experiment. The approximate location of the interaction region is indicated by the cross. Photo courtesy of CERN. 
% source: https://zh.wikipedia.org/wiki/File:ATLAS_Above.jpg
%
\textbf{(b)}
One of the four tracking stations~(TS) of ANUBIS in the $(x,z)$ plane. The shaft walls and the ATLAS cavern pipework are shown in gray, the TS in blue, the support structure of the TS in orange. The TS is made up from $1\times1~\text{m}^2$ units shown in Fig.~\ref{fig:anubis_ts_1m_unit} indicated by the squares.
}
\end{figure}

A major challenge for transverse experiments to search for LLPs with $L \gsim10~$m are the costs associated with the large active decay volume that is needed to achieve a high sensitivity exploiting large solid-angle and decay length coverage.
% combined with low backgrounds. 
The central idea of ANUBIS is to dramatically reduce the experimental costs by repurposing existing infrastructure.
This paper documents the original ANUBIS proposal exploiting the PX14 installation shaft that is not used during regular LHC operation and provides access to the ATLAS experimental cavern UX15. 
The PX14 shaft provides a large total decay volume of about 15,000~$\metre^3$ that is adjacent to the ATLAS experiment and is almost perfectly aligned towards the interaction point.
The view from the surface down the PX14 shaft is shown in Fig.~\ref{fig:shaft_above}. 
In this Figure, as in the rest of the paper, the ATLAS coordinate system\footnote{Throughout this proposal, a right-handed coordinate
  system with its origin at the nominal interaction point in the
  centre of the ATLAS detector and the $z$-axis along the beam pipe is adopted. The
  $x$-axis points to the centre of the LHC ring, and the
  $y$-axis points upward. Cylindrical coordinates $(r,\phi)$ are used
  in the transverse plane, $\phi$ being the azimuthal angle around the
  $z$-axis. 
  Transverse momentum is defined by $\pt =
  p\sin{\theta}$ where $\theta$ denotes the polar angle between the momentum and $z$-axis.} 
is adopted.

In addition to the above benefits, 
another major advantage of the PX14 shaft is that it provides an air-filled decay volume and hence features relatively low backgrounds, as elaborated in Section~\ref{sec:bkg}.
The backgrounds can be further reduced using the information about the $pp$ collision from the adjacent ATLAS detector. 
To exploit this feature, the proposal foresees a full integration of ANUBIS with ATLAS, allowing ANUBIS to trigger the readout of ATLAS. 
This is technically feasible given the ATLAS Level~0 trigger latency of 5.5~$\mu$s at the HL-LHC and the provision of an external trigger path in the central trigger processor~\cite{bib:trigtdr,bib:sankey}.
Another advantage of a full integration of ANUBIS and ATLAS is that it would provide a continuous tracking volume that extends from the interaction point to the top of the PX14 shaft, establishing sensitivity to LLP models from proper decay lengths of up to $c\tau\approx10^{6}~\metre$ down to prompt decays. 
Another benefit of the combined continuous active decay volume of ANUBIS and ATLAS is to increase the acceptance to reconstruct both LLP displaced vertices in models with pair-produced LLPs.

{\em In summary,} the existing PX14 shaft would provide sensitivity to LLP phenomena produced at the electroweak scale and above, and practically eliminate the need for underground civil engineering. 
It is particularly noteworthy that the main access shaft of the CMS detector, PX56, and the secondary ATLAS access shaft, PX16, provide similar opportunities for alternative or additional experimental setups.

To fully exploit the physics potential of the large decay volume, the ANUBIS detector concept consists of instrumenting the PX14 shaft with four Tracking Stations~(TS) spaced 18.5~m apart from each other along the shaft, providing a uniform coverage of the active decay volume. 
The geometric TS layout in the $(x,z)$ plane perpendicular to the shaft axis is shown in Fig.~\ref{fig:anubis_ts}. 
To maximize the geometric acceptance, the TS are round and concentric with the shaft cross section, except for the region $x > -7$~m to avoid the pipework shown in Fig.~\ref{fig:shaft_above}.
This setup results in a cross-sectional area of 230~m$^2$ per TS. 

\enlargethispage{5mm}
Each TS is belayed individually into the shaft from the surface using four steel ropes peripherally attached to the internal support structure of the TS. 
This technical solution requires no underground civil engineering. 
Once lowered into position, the TSs are stabilised within the shaft using cams.
The TSs are then connected to the remainder of the ANUBIS detector infrastructure that is in turn interfaced to ATLAS. 
The advantage of using cams is that the TSs can be quickly extracted after disconnecting from services in an emergency.
However,~the~TSs of ANUBIS would have to be extracted and stored at the surface during technical shutdown periods of the LHC to allow access through the PX14 shaft, and belayed back into the shaft during data taking periods.
Discussions with the ATLAS technical coordination team identified two critical points related to this~\cite{bib:ludo}.
First, while the extraction of a TS as a whole is technically feasible, it is not deemed practical because of limitations on the temporary storage space in the SX15 experimental hall above the PX14 shaft that would be blocked for the duration of the disassembly of the TS for transport for storage.
Second, the storage itself is an issue, as there is insufficient space on the Point 1 ATLAS site for four TS, and even on the main Meyrin site space is short.
Hence, other options for ANUBIS have been explored~\cite{bib:anubis_v2}.
No other critical points affecting the technical feasibility of ANUBIS were identified, neither by the ATLAS technical coordination team nor with CERN civil engineering experts.

\begin{table}
\centering
\caption{
Required performance specifications for ANUBIS. 
}
\label{tab:specs}
\begin{tabular}{lll}
\hline
\hline
Parameter & ~~& Specification \\
\hline
Time resolution && $\delta t \lsim 0.5~$ns \\
Angular resolution && $\delta \alpha \lsim 0.01$~rad \\
Spatial resolution && $\delta x,\delta z \lsim 0.5~$cm \\
Per-layer hit efficiency && $\varepsilon \gsim 98\%$ \\
\hline
\hline
\end{tabular}
\end{table}

\section{Performance specifications of ANUBIS}
\label{sec:perf}
The choice of the tracking detector technology and design discussed in Sect.~\ref{sec:technology} are determined by precise vertex resolution in space-time, which is needed to eliminate backgrounds while maintaining a high signal efficiency. The required performance specifications for ANUBIS are summarized in Table~\ref{tab:specs} and are discussed below.

\begin{itemize}
\item 
The hit time resolution should be $\delta t < 0.5$~ns or better to eliminate backgrounds from secondary particles produced in interactions of hadrons with upstream TSs, shaft walls, and cosmic rays. For $\beta=1$, $\delta t = 0.5$~ns translates into a resolution of $\delta y_\text{DV} \approx 15$~cm of the displaced vertex (DV) location along the shaft axis $y$. A precise knowledge of $y_\text{DV}$ ensures that the DV candidates are located in the fiducial region, i.e., the air-filled space between the TSs.

Moreover, $\delta t < 0.5$~ns allows the measurement of $\beta$ of charged particles traversing several TSs using the time-of-flight with a precision of down to $\delta\beta\lesssim0.001$.
\item
The resolution on the angle $\alpha$ between a particle track and the $y$ axis defines the reach of ANUBIS at low LLP masses \mllp. This is a purely kinematic requirement: the opening angle $\omega$ between particles from a two-body LLP decay is inversely proportional to their average boost $\frac12\cdot\frac{\meff}{\mllp}$, where \meff is the mass of the mediator decaying into a pair of LLPs. Thus, $\mllp \approx \frac 12\meff\cdot \omega$. Assuming that the LLPs are produced around the electroweak scale with $\meff \approx 100~\GeV$ and considering $\delta\omega \approx \sqrt 2\cdot \delta\alpha$ for a symmetric LLP decay into a fermion-antifermion pair, a precision of $\delta \alpha \lsim 0.01$~rad would allow for sensitivity to $\mllp \gsim 0.5~\GeV$, i.e., just above the $K_L$ mass. For smaller $\delta\alpha$, ANUBIS gains sensitivity to smaller \mllp values. 
The assumption of a two-body decay is conservative, as DVs with higher charged particle multiplicities are experimentally easier to identify.
\item
The spatial resolution in the TS plane $(x,z)$ is directly connected to the required angular resolution. Assuming two hits and a lever arm of 1~m corresponding to the TS depth, $\delta\alpha\lsim0.01$~rad translates into $\delta x=\delta z\lsim 2\delta\alpha \times 1~\text{m}= 5~$mm. Smaller values of $\delta x$ and $\delta z$ result in increased sensitivities to smaller \mllp.
\item
The efficiency of capturing a signal from potential LLP decays and of rendering SM backgrounds negligible is defined by the efficiency to identify hits. To achieve a per-TS efficiency of 95\% or better, a per-layer hit efficiency of $\varepsilon\gsim 98\%$ is required assuming two layers per TS.
\end{itemize}

\begin{figure}[tb]
\centering
\begin{subfigure}{0.49\textwidth}
\includegraphics[height=0.8\textwidth]{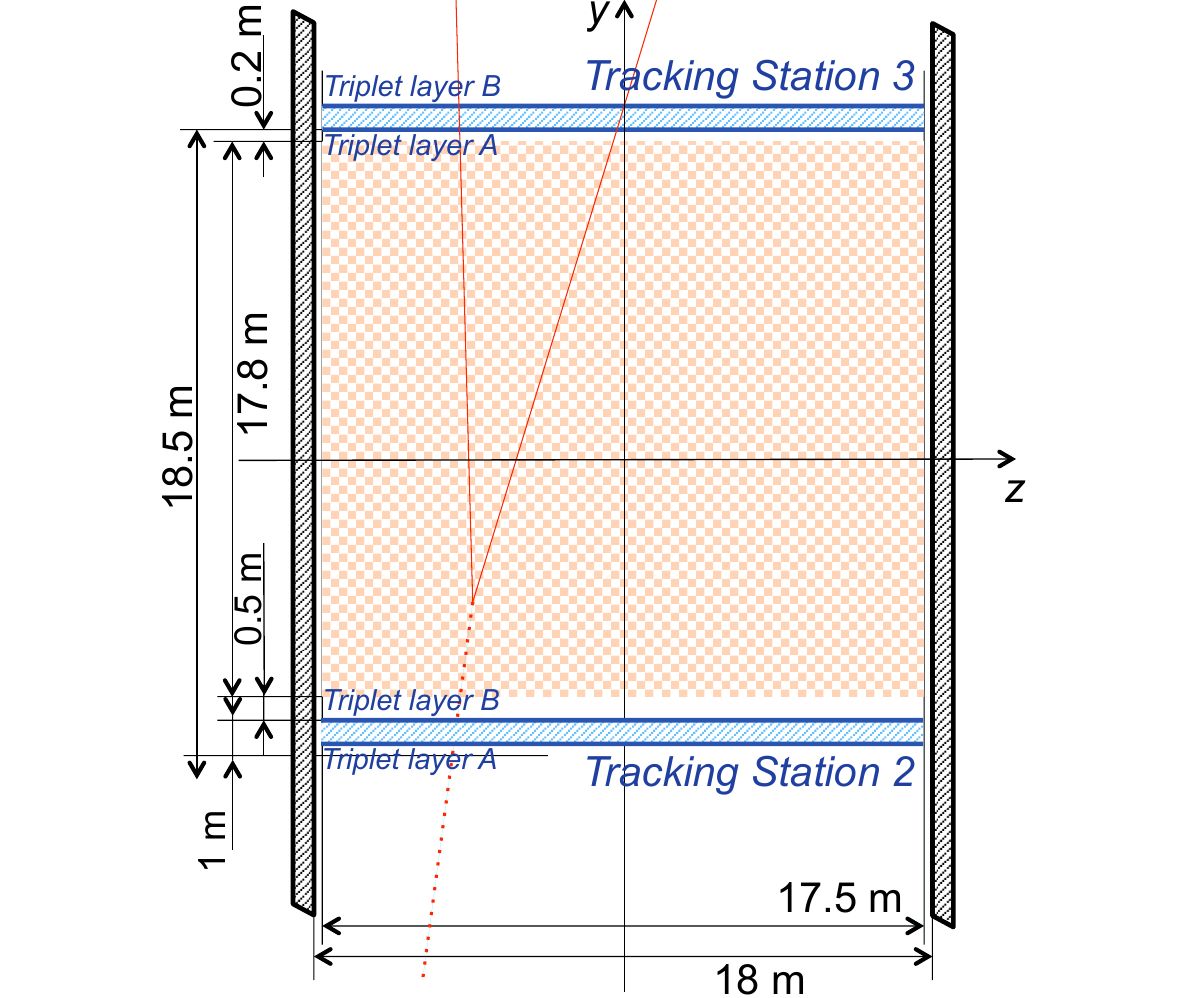}
\caption{}
\label{fig:anubis_yz}
\end{subfigure}
\begin{subfigure}{0.49\textwidth}
\includegraphics[height=0.8\textwidth]{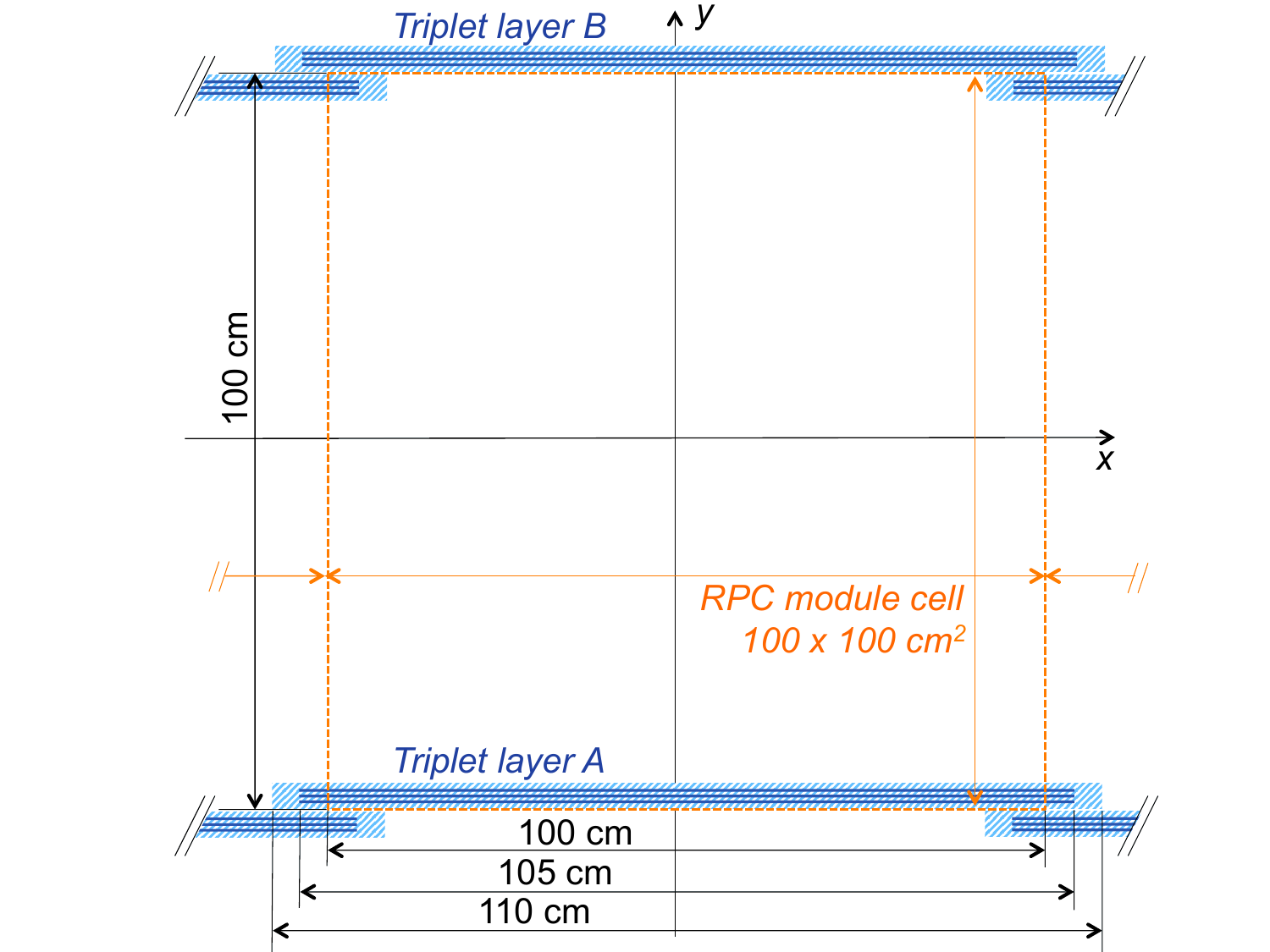}
\caption{}
\label{fig:anubis_ts_1m_unit}
\end{subfigure}
\caption{
\label{fig:ts_sketch}
\textbf{(a)}
Cross section of the PX14 ATLAS shaft in the $(y,z)$ plane with two representative TSs. The shaft walls are indicated in hatched gray, the TSs in shaded blue, and the fiducial region in checkered orange. The layers of both TSs are shown in solid blue, and a potential LLP decay into a pair of muons is sketched in red.
\textbf{(b)}
Side-view in the $(x,y)$ plane of one of the $1\times1~\text{m}^2$ RPC module cell units composing the ANUBIS tracking stations. The unit consists of two layers A and B (shaded blue) with a triplet of detection layers each.
}
\end{figure}

The performance specifications discussed above can be met if each TS nominally consists of two layers of tracking detectors spaced one meter apart along the $y$ axis, providing a lever arm for track reconstruction that is one meter or longer. A cross-section of the shaft in the $(y,z)$ plane with two representative TSs is shown in Fig.~\ref{fig:anubis_yz}. 

The calibration of the detector's spatial coordinate and time measurements that is necessary to achieve the performance specification in Table~\ref{tab:specs} can be performed using the copious source of muons from LHC collisions as well as muons from cosmic rays.

\section{Backgrounds}
\label{sec:bkg}

The ANUBIS experiment is expected to be close to background-free. 
The potential background sources are discussed below.

Backgrounds from cosmic rays can be effectively vetoed using timing and directionality requirements assuming the performance specifications in Table~\ref{tab:specs}. 
The distance of 1~m between the tracking layers of an ANUBIS tracking station allows to discriminate downwards-going from upwards-going cosmic ray particle tracks at a level of 7$\sigma$ assuming a timing resolution of $\delta t=0.5$~ns. 
The incidence rate of upwards-going tracks from cosmic ray scattering is negligible given the low probability of scattering with the air-filled active volume of ANUBIS, combined with the very low probability of obtaining more than one upwards-going track to form a candidate vertex.
Moreover, the decreasing flux of cosmic ray particles inside the shaft with increasing distance from the surface will provide a utile handle to verify the absence of any notable background contribution from this source.

Beam-induced backgrounds like beam-gas and beam-collimator interactions are negligible for ANUBIS given its position: the active volume starts several metres away from the beamline in the radial direction. 
Furthermore, the topology of beam-induced backgrounds is strongly boosted along $z$ (the beamline) and hence very different from the topology of DVs from LLPs originating from the interaction point, which provides a handle to reduce this background.
Moreover, the difference in topology provides an additional handle to reject beam-induced backgrounds using  timing information.

Backgrounds from decays of quasi-thermal neutrons~\cite{Hedberg:2004lua} are negligible since the momenta of their decay products are only about 1~MeV, which is insufficient to generate signals in both the upper and lower layers of a tracking station and hence will not be reconstructed as a track, let alone a DV.

Instrumental backgrounds can arise from random coincidences of unrelated particle tracks so that they are reconstructed as one DV.
The contribution of this type of background is inversely proportional to the density of tracks in space-time.
It is therefore not surprising that this source of background was found to contribute only in searches for DV production using the inner detector, i.e., for decay lengths $L$ of $\mathcal{O}(10~\rm cm)$~\cite{ATLAS:hdmi}, and to be negligible for larger $L$.
Hence, this source of background is negligible for ANUBIS that will consider DVs reconstructed several metres away from the interaction point. %only %relevant in situations with high density of tracks that 

The dominant contribution to backgrounds in LLP searches typically stems from SM particles with macroscopic but not infinite lifetimes, which will be termed SM LLPs in the following. % and hence represent an irreducible background.
Among those SM LLPs, the only ones that have a sufficiently long lifetime to potentially provide a background contribution to ANUBIS are long-lived neutral kaons $K_L$ and neutrons $n$ with mean proper lifetimes of $c\tau=15.3$~m and $2.7\times10^{11}$~m, respectively~\cite{bib:pdg2018}. 
There are two main mechanisms for SM LLPs to contribute: via decays into charged particles that are reconstructed as a DV and via hadronic interactions, i.e., the production of hadrons in collisions of SM LLPs with nuclei as they propagate through matter.

Given their large $c\tau$ and typical boost factors of $\gamma\gtrsim20$, only a small fraction of $K_L$ decay inside the ANUBIS active volume. %assuming gamma=20, the fraction is 20%
Moreover, the $K_L$ decay into at most two tracks from charged particles with the exception of a few rare decay modes of which the largest one, $K_L\to \pi^\pm e^\mp\nu e^+e^-$, contributes a branching ratio of $1.2\times10^{-5}$~\cite{bib:pdg2018}. 
The DVs from $K_L$ decays are characterised by a small opening angle of a couple of degrees on account of the small $K_L$ mass, which are easy to separate from a potential LLP signal. % opening angle 2m/pT=5 deg for 10 GeV K_L
The decays of $n$ are negligible inside the ANUBIS active volume and are not considered further.

The dominant mechanism for $K_L$ and $n$ to contribute as backgrounds to ANUBIS are hadronic interactions, where typically a handful of charged particle tracks with a sizeable opening angle are produced. % in such hadronic interactions.
This signature makes hadronic interactions challenging to separate from true LLP decays.
The contribution from hadronic interactions is linearly proportional to the density of matter traversed by SM LLPs and depends on its composition. 
The probability for a hadronic interaction to occur is quantified in units of the nuclear interaction length \LI.
Particles traversing ATLAS on their way to ANUBIS will typically propagate through detector material like the ATLAS calorimeters ($\LIx{Tile}=30.3~\cm$), steel support structures ($\LIx{Fe}=16.8~\cm$), ATLAS muon spectrometer ($\LIx{MS}=385~\cm$, obtained by averaging contributions from steel and aluminium~\cite{ATLAS:2020ell} within the volume defined according to Eq.~\eqref{eq:leff_atlas} below), and air ($\LIx{Air}=7.5\times10^{4}~\cm$)~\cite{bib:pdg2018}.
Hence, it is $\mathcal{O}(10^4)$ times less likely for a background event from hadronic interaction to occur in air than in detector material.
Therefore, the {\em air-filled} decay volume of ANUBIS represents its biggest advantage, combined with its large size of 15,000~m$^3$ and proximity to the interaction point.
The position of ANUBIS behind the ATLAS detector relative to the interaction point represents another crucial advantage: the flux of neutral hadrons within the ANUBIS acceptance is dramatically reduced by the ATLAS detector, primarily by its calorimeters that account for $9.7~\LI$ at $\eta=0$ and even more with increasing $|\eta|$~\cite{Aad:2008zzm}.
This results in an attenuation of the SM LLP flux by a factor of at least $1.6\times10^{4}$ on account of the calorimeters alone and discounting any support structures, the inner detector, and the solenoid magnet that contribute several \LI of shielding in addition.

The attenuation of the SM LLP flux by ATLAS by several orders of magnitude is not yet sufficient to eliminate it completely given the total inelastic cross section of about 80~mb that is dominated by QCD multijet processes~\cite{bib:xsectotem,bib:xsecalfa,bib:xseclhcb}. 
However, this source of background can be significantly reduced by exploiting the fact that hadrons like $K_L$ and $n$ are produced inside jets, i.e., in close angular proximity $\dR\equiv\sqrt{\Delta\eta^2+\Delta\phi^2}$ to prompt jets or energetic tracks from other charged hadrons.
In practice, this is implemented as a veto on events with candidate LLP DVs that are closer than $\dR(\text{DV,jet})<0.5$ to prompt energetic jets reconstructed by ATLAS with a minimum transverse momentum $\pt>15~\GeV$.
In addition, events with candidate LLP DVs with $\dR(\text{DV,track})<0.5$ relative to any prompt energetic charged particle track reconstructed by ATLAS with a minimum transverse momentum $\pt>5~\GeV$ are vetoed to eliminate background contributions from SM LLPs inside jets with few high-momentum particles.
Finally, a missing transverse momentum $\met>30~\GeV$ reconstructed by ATLAS is required exploiting the fact that LLPs that decay inside the ANUBIS active volume and hence outside of ATLAS contribute to the detected transverse momentum imbalance.
The above selection requirements represent an {\em active} ATLAS veto that uses information about the candidate $pp$ collision event and highlights the need to synchronously trigger and read out both ANUBIS and ATLAS detectors.

The selection requirements of the active ATLAS veto are inspired by the ATLAS search for LLPs with decay lengths of $\mathcal{O}(3~\text{m})$ that produce DVs reconstructed by the ATLAS muon spectrometer~\cite{Aaboud:2018aqj}.
This makes it very similar to future searches using the ANUBIS detector, albeit with a somewhat shorter decay length. 
This ATLAS search analysed 36~\fb of $pp$ collisions at $\sqrt s = 13$~TeV. %, which is very close to $\sqrt s=14$~TeV of the HL-LHC.
This ATLAS search is used to estimate the backgrounds for ANUBIS since it is very similar in terms of the selection requirements, targeted models, and $\sqrt s$. 
The DV+$\met$ analysis strategy is followed considering the barrel region, i.e., $|\eta|<0.7$, which is most similar to ANUBIS both in terms of kinematics and instrumentation.
As for most LLP searches, the background estimate for this analysis is very difficult to model through MC simulations. 
Therefore, it is derived using a data-driven approach using uncorrelated sidebands also known as the ABCD method and reads $243\pm38~\stat\pm29~\syst$, which is consistent with the measured value $N_{\rm bgr,ATLAS}=224$.

The background estimate for ANUBIS $N_{\rm bgr, ANUBIS}$ is obtained by scaling $N_{\rm bgr,ATLAS}$ under the conservative assumption that a potential signal contribution to $N_{\rm bgr,ATLAS}$ is negligible at the precision level of the ATLAS search using 36~\fb of $pp$ collisions.
The scaling is performed according to:
\begin{equation}
\label{eq:bg}
N_{\rm bgr, ANUBIS} 
= \frac{ \mathcal{L} }{ \mathcal{L'} }
\times \frac{ \LIx{MS} }{ \LIx{Air} }
\times \frac{\varepsilon}{\varepsilon'} 
\times \frac{ \leff }{ \leff' }
\times f_{\rm bgr,MS}
\times N_{\rm bgr, ATLAS}\,,
\end{equation}
where the primed quantities refer to ATLAS and non-primed to ANUBIS, $\mathcal{L}$ is the integrated luminosity, $\varepsilon$ denotes the reconstruction efficiency, and $f_{\rm bgr,MS}$ represents the fraction of DVs originating within the ATLAS muon spectrometer effective volume.
The geometric differences between the active volumes $V$ of ATLAS and ANUBIS are captured through the effective path length  $\leff\equiv\int_{\rm V} \dif\Omega \dif\ell$, i.e., the path element $\dif\ell$ and the solid angle element $ \dif\Omega$ integrated over $V$, since the angular flux $\frac{ \dif N}{\dif\Omega dt} = \const$.
The calculation of \leff conservatively ignores the decrease of the SM LLP flux with $\ell$ as it is depleted by hadronic interactions.

For ANUBIS, the effective path length is calculated following
\begin{eqnarray}
\leff&=&\int_{V=\rm shaft+cone} \dif\Omega \dif\ell \nonumber\\
&=& \int_{V=\rm shaft+cone} \frac 1{|\vec r|^2} \dif^3\vec r\nonumber\\
&=& 11.3~\metre\,,
\label{eq:leff_anubis}
\end{eqnarray}
where conservatively the volume of the shaft has been used in full, i.e., including the non-projective sections, and the cone pointing from the interaction point to the lowest TS with the cone peak capped at a radial distance of 7~m from the beamline.

For ATLAS, the effective path length is determined according to
\begin{eqnarray}
\leff'&=&\int_{V'=\rm ATLAS} \dif\Omega \dif\ell \nonumber \\
&=& \int_{0}^{2\pi}\int_{0.92}^{2.22}\int_{R_{\rm in}}^{R_{\rm out}} \dif\phi \dif\theta\sin\theta \dif\ell \nonumber\\
&=& 22.1~\metre\,,
\label{eq:leff_atlas}
\end{eqnarray}
where the azimuthal boundaries reflect the $|\eta|<0.7$ requirement, while the radial boundaries ${R_{\rm in}}=4.3$~m and $R_{\rm out}=7$~m correspond to the outer edge of the calorimeter and the dramatic drop in the DV reconstruction efficiency of the ATLAS muon spectrometer~\cite{bib:dvalgo}, respectively.

The fraction of DVs originating within the ATLAS muon spectrometer effective volume is estimated at $f_{\rm bgr,MS}=0.65$ by  calculating the average DV density as a function of radial distance from the beamline. 
The migration of vertices from the calorimeter into the MS due to the finite resolution of the DV reconstruction algorithm is accounted for under the assumption that the DV density is proportional to the average material density.

With the conservative estimates of the effective path length from Eqs.~\eqref{eq:leff_anubis} and \eqref{eq:leff_atlas}, using a projected HL-LHC integrated luminosity of 3~ab$^{-1}$, and conservatively assuming that the ANUBIS detectors will be two times more efficient than the current ATLAS muon system, Eq.~\eqref{eq:bg} yields:
\begin{eqnarray}
N_{\rm bgr, ANUBIS} 
&=& \frac{ 3\,\iab }{ 36\,\ifb }
\times \frac{ 385\,\cm }{ 7.50\times10^4\,\cm }
\times 2 \nonumber 
%&\times& \frac{ \int_{V=\rm ANUBIS} \dif\Omega \dif\ell }{ \int_{V'=\rm ATLAS} \dif\Omega \dif\ell }
\times \frac{ \leff }{ \leff' }
\times f_{\rm bgr,MS}
\times N_{\rm bgr, ATLAS} \nonumber \\
&=& 0.856 
\times \frac{ 11.3~\text{m} }{ 22.1~\text{m} }
\times 0.65
\times N_{\rm bgr, ATLAS} \nonumber \\
&=& 0.284%2 
\times N_{\rm bgr, ATLAS} \nonumber \\
&=& 63.7 \pm 4.3\,.\label{eq:bgnum}
%&=& 63.7\,.\label{eq:bgnum}
\end{eqnarray}
This background level allows to assert the observation of a potential signal at 5$\sigma$ significance with $N_{\rm sig}\geq52$ detected signal events~\cite{Cowan:2010js}.
$CLs$ exclusion limits at 95\% confidence level (CL) can be set with $N_{\rm sig}=19$ events.
Sensitivity projections in Section~\ref{sec:reach} assume that the observation of $N_{\rm sig}\geq50$ signal events would be sufficient for a BSM LLP signal discovery.
These projections will be referred to as `conservative' in the following.

Several conservative assumptions have been made throughout the estimate above.
Therefore, the number of background events is expected to be substantially lower than the data-driven projection of Eq.~\ref{eq:bgnum}.
Hence, the sensitivity estimate in Section~\ref{sec:reach} provides a second `background-free' projection that requires $N_{\rm sig}\geq4$ for an exclusion at 95\% CL. 

The true sensitivity of the ANUBIS experiment is bounded by the `background-free' and the `conservative' projection, where the latter is a conservative data-driven estimate.
It should be added that the search becomes essentially background-free as soon as a second DV is found in the active volume of either the ANUBIS or ATLAS detector, or a prompt object like e.g., a $Z$ boson, is registered by the ATLAS detector.

%%%%%%%%%%%%%%%%%%%%%%%%%%%%
\section{The ANUBIS Detector Technology and Cost Estimate}\label{sec:technology}
%%%%%%%%%%%%%%%%%%%%%%%%%%%%

An ANUBIS detector configuration featuring four TSs provides an acceptable dead area due to the slightly non-projective orientation of the PX14 shaft relative to the interaction point, and is therefore a good compromise between physics performance and cost
Each tracking station has an area of $230~\sqm$ and features two detector layers that are one metre apart. 
Assuming $1\times1~\sqm$ unit detector cells with a 5~cm stashed overlap on each side as shown in Fig.~\ref{fig:anubis_ts_1m_unit}, this translates in a total instrumented area of $2\times290~\sqm=580~\sqm$ per TS.
In turn, this results in a total instrumented area is about $2.3\times10^{3}~\sqm$ for the original ANUBIS proposal.

The large instrumented area of ANUBIS calls for a cost-effective detector solution, which, at the same time, is able to meet the performance requirements set out in Section~\ref{sec:perf} and summarised in Table~\ref{tab:specs} in order to meet ANUBIS' physics goals.
A survey of detector technologies identified Resistive Plate Chambers (RPC) as the preferred detector technology for ANUBIS.
In particular, the RPCs from the Phase II upgrade of the ATLAS muon system that were designed to cope with unprecedented instantaneous luminosity at the HL-LHC were found to be the preferred choice~\cite{bib:muontdr}.
This choice was made following physics performance, modest overall costs, and practical considerations.
The technology is mature: very similar RPCs were used for the Phase I upgrade of the BIS7 sector of the ATLAS muon system installed in 2020, drawing a clear R\&D roadmap towards a full scale ANUBIS detector.
%In particular, the BIS7 RPCs from the ATLAS Phase I upgrade to be installed at ATLAS in 2020 were found to be the preferred choice~\cite{bib:muontdr}. 

The heart of the Phase~I BIS7 RPC~(and of the Phase~II upgrade RPC) is a 1~mm wide gas gap~\cite{bib:muontdr} that provides a time resolution of 350 ps per RPC detector singlet, which is further improved to 200~ps for a triplet of RPC detectors that will form a unit cell -- the basic building block of a TS layer.
Despite the narrow 1~mm gas gap, the BIS7 RPCs are highly efficient due to a new generation amplifier implemented ASIC in Silicium-Germanium (SiGe) technology that has a signal threshold of only about a fC and a rate capability of 10~kHz/cm$^2$~\cite{bib:muontdr}.
An RPC singlet can provide an efficiency of $\varepsilon>98\%$~\cite{bib:rpcperf2}, which translates into an even higher efficiency for an RPC triplet using a two-out-of-three hit reconstruction logic.
A spatial resolution of a fraction of a cm for multi-strip hits can be achieved using analogue readout~\cite{,bib:rpcperf3}.
Besides meeting the physics performance requirements from Table~\ref{tab:specs} at a reasonable price, the BIS7 RPCs have another crucial advantage: the technology is mature, minimising the amount of additional detector R\&D, which in turn allows to further compress the timeline to finalise the detector design and to minimise the overall cost.

Considering the production yield, BIS78 RPCs using components from commercial suppliers cost about 2.5~kCHF/\sqm~\cite{bib:aielli}, including the gas gap, readout strip panels, and frontend readout.
The tracking detectors represent by far the dominant cost item, and the total costs of ANUBIS including minor civil engineering are estimated at $\mathcal{O}(10~\text{MCHF})$ in a configuration with four TSs. 
The weight of the RPC triplet modules including detector cage mechanics is 30~kg/\sqm~\cite{bib:aielli}, resulting in about 20~tonnes per TS. This is well manageable with the existing infrastructure, even if allowing for a very generous factor of five for mechanical support and other infrastructure. 
Alternative detector technologies like finely granulated scintillators have been considered but were found to not provide a reduction of costs while meeting the physics performance specifications from Table~\ref{tab:specs}.

%%%%%%%%%%%%%%%%%%%%%%%%%%%%
\section{Sensitivity of ANUBIS}% 
\label{sec:reach}
%%%%%%%%%%%%%%%%%%%%%%%%%%%%
To estimate the reach of ANUBIS, a standard benchmark scenario for LLP searches at the LHC is employed: exotic Higgs boson decays to new long-lived scalars $h \to ss$~\cite{bib:htoss}, inspired by Neutral Naturalness~\cite{Chacko:2015fbc,Burdman:2006tz,Cai:2008au,Chacko:2005pe,Argyropoulos:2021sav}.
This model represents a perfect scenario to highlight the complementarity between different types of experiments: 
it cannot be probed at forward experiments like FASER and beam-dump experiments like SHiP, as it requires LLP production at the electroweak scale with $\sqrt{\hat s} > m_h$ where $m_h$ is the mass of the Higgs boson~\cite{Antel:2023hkf}. 
While this model is kinematically accessible at the main LHC experiments like ATLAS and CMS (and to some extent LHCb), for decay lengths of $\mathcal{O}(10~\metre)$ it can only be probed at transverse experiments like ANUBIS.
It is hence not surprising that this benchmark scenario is highlighted in the recent ESPPU submission of the Physics Beyond Colliders initiative~\cite{PBC:2025sny}.

The $h \to ss$ benchmark process is generated using the \texttt{MadGraph5} \cite{Alwall:2014hca} MC simulation programme. 
For the sensitivity estimate, the number of particle trajectories that enter the active volume of ANUBIS and decay before they penetrate at least one TS is counted. 
This study considers a geometry with four TSs that are assumed to be fully efficient, which is a good approximation given that a triplet of RPC detectors has $\varepsilon>98\%$.
Two configurations are examined: 
in the `shaft-only' configuration only LLP decays in the PX14 shaft are considered, whereas in the `shaft+cone' configuration the LLPs are allowed to decay inside the shaft and in the cone between the interaction point and the lowest TS at the bottom of the PX14 shaft, where the cone tip is capped at the boundary of the muon spectrometer of ATLAS.
For each of the above configurations two estimates motivated in Section~\ref{sec:bkg} are provided: a `background-free' and a  `conservative' estimate requiring $N_\text{sig}\geq4$ and 50 for the observation of the $h \to ss$ decays, respectively.
This results in a total of four sensitivity projections.

\begin{figure}[h]
\begin{center}
\includegraphics[width=0.59\textwidth]{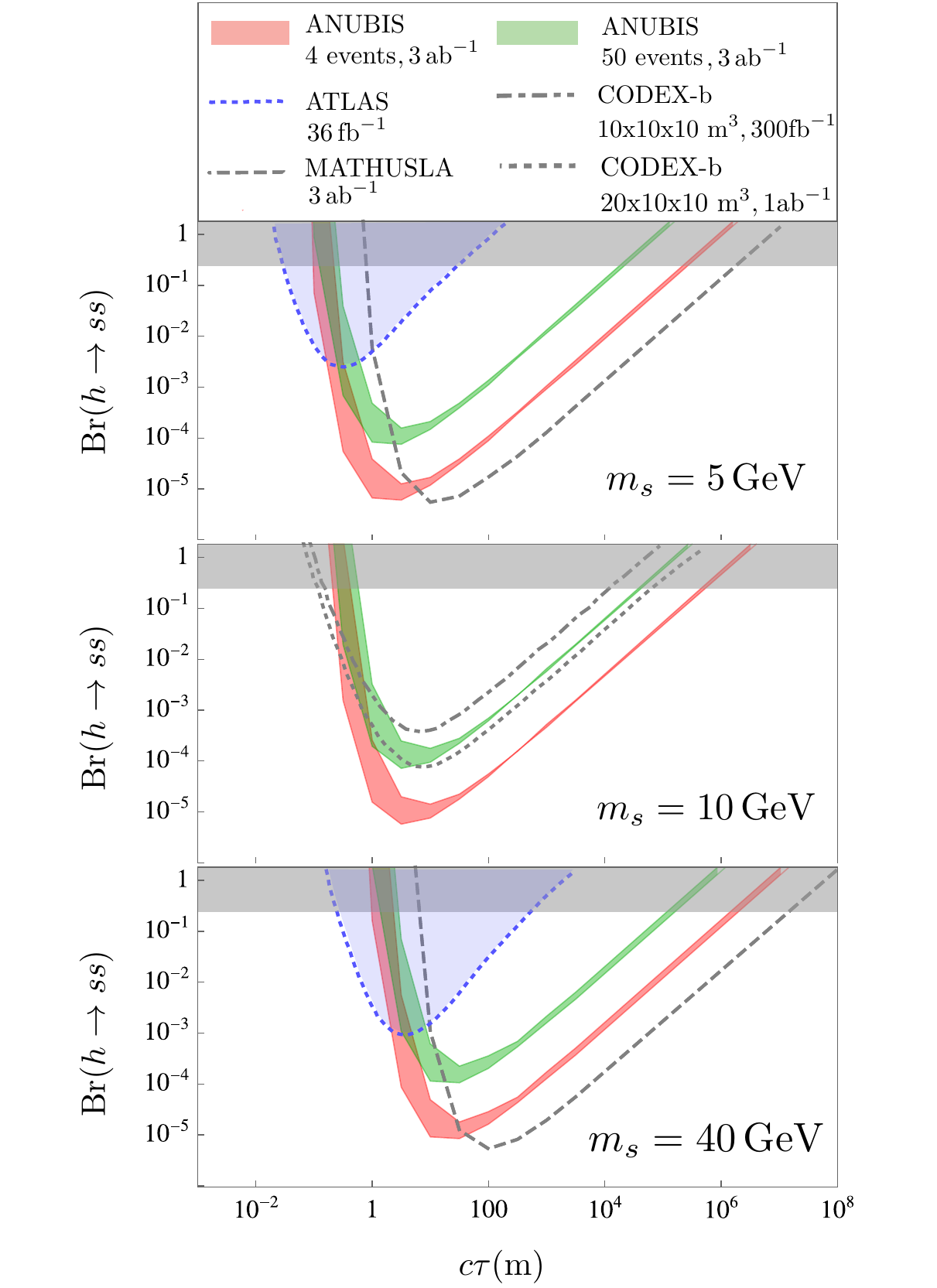}
\end{center}
\vspace{-.5cm}
\caption{
Projected sensitivity of ANUBIS, CODEX-b~\cite{Gligorov:2017nwh}, and MATHUSLA~\cite{Chou:2016lxi} for LLPs for Higgs decays $h \to ss$ with $m_s=5, 10$ and $40$ GeV at the HL-LHC with $\sqrt s = 14$~TeV. %  with $\mathcal{L}=3$ ab$^{-1}$. 
The projection bands `ANUBIS~4~events' and `ANUBIS~50~events' are given for the `background-free' and `conservative' background cases requiring 4 and 50 signal events, respectively, as motivated in Section~\ref{sec:bkg}. 
The upper~(lower) boundary of the projection bands corresponds to the `shaft-only' (`shaft+cone') configurations defined in the text. 
%impact of extending ANUBIS' active decay volume up to the ATLAS muon spectrometer is indicated by the filled area. 
The grey parameter space is excluded by the existing constraints on invisible Higgs decays~\cite{Sirunyan:2018owy,Aaboud:2019rtt}. 
The region excluded by a dedicated Higgs portal LLP search with ATLAS is shown for reference~\cite{Aaboud:2018aqj}.
}
\label{fig:sensitivity}
\end{figure}

The projected sensitivity reach of the original ANUBIS proposal using 3~ab$^{-1}$ of integrated luminosity is shown Fig.~\ref{fig:sensitivity} for LLP masses of $m_s=5, 10$ and $40$ GeV. 
The `background-free' case requiring $N_\text{sig}\geq4$ is represented by the filled red band, where the upper~(lower) boundary corresponds to the `shaft-only'~(`shaft+cone') configuration. 
The maximum sensitivity is reached for $c\tau$ between 1~m for $m_s=5~\GeV$ and 100~m for $m_s=40~\GeV$, allowing to probe $h\to ss$ branching ratios down to $\br(h\to ss)=10^{-5}$.
The expected HL-LHC bounds of 2.5\% on $h\to ss$ from $h\to(\text{invisible})$ decays~\cite{bib:hinv_hllhc} are surpassed up to $c\tau=10^{5}$~m~($10^6$~m) for $m_s=5~\GeV$~(40~GeV).
The `conservative' background case requiring  $N_\text{sig}\geq50$ is shown by the filled green band, with the `shaft-only'~(`shaft+cone') configuration represented by the upper~(lower) boundary of the band.
This tightened $N_\text{sig}\geq50$ requirement of the `conservative' case increases the $Br(h\to ss)$ evidence threshold by only one order of magnitude, which still allows to probe $h\to ss$ branching ratios down to $\br(h\to ss)=10^{-4}$.

ANUBIS' sensitivity in Fig.~\ref{fig:sensitivity} is compared to sensitivity projections from other proposals that were available in 2019, when the original ANUBIS proposal was made: 
the $10\times10\times10~$m with 300~$\ifb$ and $20\times10\times10~$m with 1~ab$^{-1}$ scenarios of \hbox{CODEX-b}~\cite{Gligorov:2017nwh} as well as 
the $200\times 200\times 20~\text{m}^3$ version of MATHUSLA~\cite{Chou:2016lxi} using 3~ab$^{-1}$. 
For reference, the limit set by ATLAS that provides the background estimate discussed in Section~\ref{sec:bkg} is also shown~\cite{Aaboud:2018aqj}. 
The sensitivity of the original ANUBIS proposal surpasses the standard CODEX-b scenario in the entire relevant phase space, outperforms MATHUSLA for LLPs with $c \tau \lesssim 10\,$m, and extends the reach of ATLAS by many orders of magnitude at $c \tau \gtrsim 1\,$m.

%%%%%%%%%%%%%%%%%%%%%%%%%%%%
\section{Conclusion}
\label{sec:summary}
%%%%%%%%%%%%%%%%%%%%%%%%%%%%

Long-lived particles arise naturally in many extensions of the SM, including scenarios with dark matter candidates and models capable of explaining small but non-zero neutrino masses, the origin of the matter-antimatter asymmetry, etc.  
This paper presents a yet unexplored and unique opportunity to close the present gap in sensitivity for LLPs with decay lengths $L$ of $\mathcal{O}(10~\metre)$ produced at the electroweak scale or above, i.e., $\sqrt{\hat s}\gtrsim80~\GeV$, which cannot be probed by any of the existing or approved future experiments. %for  $m_\text{LLP}>1~\GeV$, 
This can be achieved at a modest cost by instrumenting the existing air-filled infrastructure around the ATLAS or CMS experiments, e.g., the PX14 service shaft of ATLAS,  with a series of tracking stations. 

The original proposal for the ANUBIS detector was outlined, describing its main features, specifying its performance requirements, proposing a detector technology with preliminary cost estimates, and demonstrating its unique projected sensitivity. 
The original ANUBIS proposal is expected to substantially extend the sensitivity of ATLAS and CMS, surpass CODEX-b, and provide a sensitivity similar to that of MATHUSLA. However, the total instrumented area of ANUBIS of 2,320~m$^2$ corresponds to less than 10\% of MATHUSLA's, which is expected to be reflected in the costs.

It should be stressed that, even though this proposal was made specifically for the PX14 ATLAS shaft, the idea of ANUBIS is equally applicable to both ATLAS and CMS, and in fact to any large air-filled potential active volume in close proximity to a general purpose detector that can serve as an active veto. 
Meanwhile, an alternative ANUBIS proposal was provided that foresees instrumenting the ceiling of the ATLAS UX15 cavern and the bottom of the PX14 and PX16 service shafts of ATLAS, which will be documented in a future paper.
%and using the upper part of the ATLAS $\mathcal{O}$

%Alternative detector technologies, such as finely granulated scintillators, can be used and may well further reduce the total costs of ANUBIS relative to this proposal.

To demonstrate the feasibility of the ANUBIS project, the construction of two $1\times1$~m$^2$ tracking station unit prototypes to be suspended at the bottom and top of the PX14 shaft for Run~3 of the LHC was proposed. 
Meanwhile, a prototype called proANUBIS was constructed using $1\times1.8~\metre^2$ Phase I RPCs for the BIS sector~\cite{bib:muontdr}, and has been taken physics data since 2024~\cite{Revering:20247,Shah:2024uj}.
The proANUBIS demonstrator has been invaluable to better understand the detector and readout implementations, mechanical engineering needs, and to further refine \textsc{GEANT4}-based background simulations for ANUBIS~\cite{Geant4}.

In summary, the unique opportunity of large, pre-excavated spaces close to the ATLAS and CMS interaction points must be explored in order to fully exploit the discovery potential of the HL-LHC -- the only apparatus capable of copiously producing new particles above the electroweak scale in a laboratory. No discovery opportunities should be left unexplored.

\section*{Acknowledgments}
The authors would like to thank the U. of Naples particle physics group for their hospitality during the ATLAS Exotics workshop where key parameters of this idea were fleshed out. 
The authors would also like to thank Giulio Aielli, Jamie Boyd, Chip Brock, Andrea Coccaro,  Tova Holmes, Umberto De Sanctis, Lars Henkelmann, Felix Kling, Christopher Lester, Ludovico Pontecorvo, Jianming Qian, Michael Revering, Patrick Rieck, Aashaq Shah, and Paul Swallow for the fruitful discussions and/or the help with the cost estimate. 
The author would further like to express their sincere gratitude to the ATLAS Collaboration and the ATLAS Technical Coordination team for their support with the installation and operation of the proANUBIS demonstrator. 
LL is supported by the US Department of Energy under Grant DE-SC0007881. 
CO is supported by the Swedish Research Council under Grant 2017-05160.

\clearpage

%\bibliography{mf}

%\hbox{}
\printbibliography[title=References,heading=bibintoc]

\clearpage
%\onecolumn
% %%%%%%%%%%%%%%%%%%%%%%%%%%%%
\section*{Supplementary material}
\label{App}

\begin{figure*}[h]
\begin{center}
\includegraphics[width=0.5\textwidth]{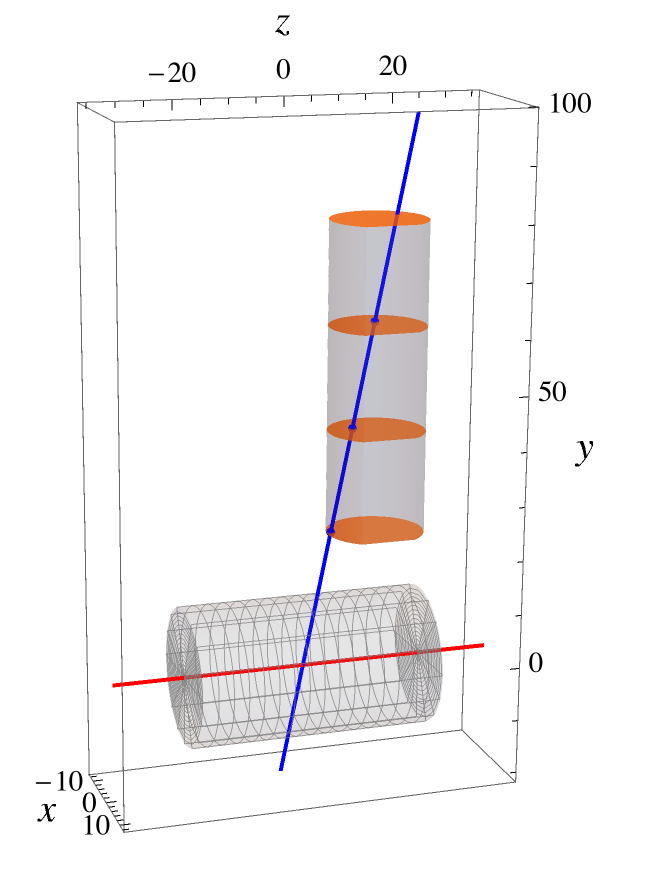}
\end{center}
\caption{
Geometry of the proposed ANUBIS tracking stations (shown in orange) in the ATLAS installation pit together with an example LLP trajectory from a Higgs decay that would be reconstructed if it decays in the detector volume (blue) and the ATLAS detector with the beam axis shown in red.
}
\label{fig:anubis_geometry}
\end{figure*}
\begin{figure*}
\begin{center}
\includegraphics[width=0.6\textwidth]{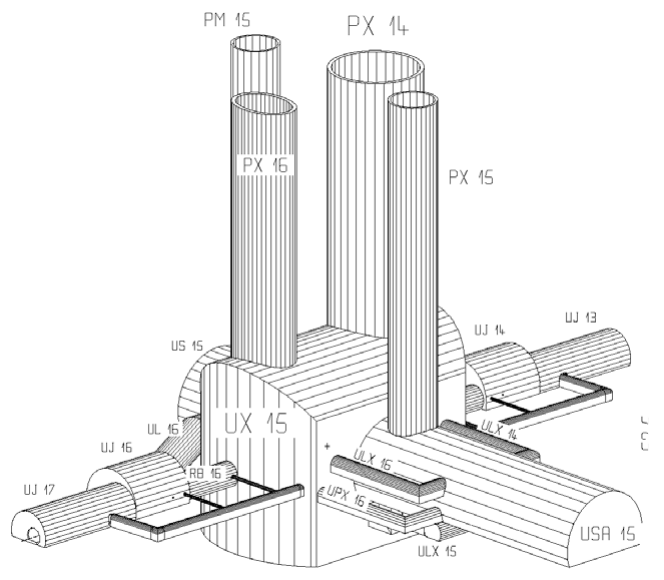}
\end{center}
\caption{
Layout of the UX15 ATLAS experimental cavern with access shafts~\cite{Dawson:2004pta}. 
The centre of the LHC ring is to the left of the cavern.
}
%\label{fig:fluence}
\end{figure*}
\begin{figure*}
\begin{center}
\includegraphics[width=0.8\textwidth]{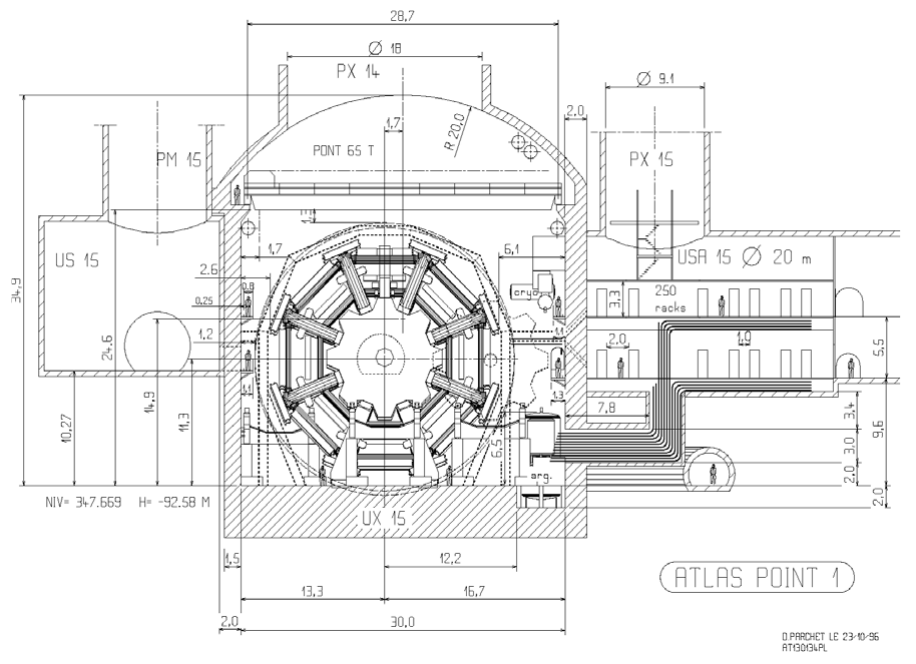}
\end{center}
\caption{
Cross-section of the UX15 ATLAS experimental cavern~(Point~1)~\cite{Dawson:2004pta}. 
The centre of the LHC ring is on the left in this drawing.
Note that the PX14 shaft is exactly centered on ATLAS.
}
%\label{fig:fluence}
\end{figure*}

\end{document}